# Modelando el espectro de potencia de la encuesta espectroscópica de galaxias eBOSS

Dante V. Gomez-Navarro[1]

[1]Instituto de Física, Universidad Nacional Autónoma de México, Ciudad Universitaria, CP 04510, México D.F., México
E-mail: dantegomezn@gmail.com

*Resumen* — **Actualmente existe mucho interés en extraer información cosmológica de las encuestas de galaxias espectroscópicas ya que permiten obtener una medición directa de la tasa de crecimiento de las estructuras y la historia de expansión. Motivados por este hecho, en este trabajo modelamos los multipolos del espectro de potencia de la encuesta de galaxias eBOSS (extended Baryon Oscillation Spectroscopic Survey) usando la teoría de perturbaciones no lineal.**

*Palabras Clave* – **Encuestas de galaxias espectroscópicas, estructuras a gran escala del Universo, teoría de perturbaciones no lineal**

*Abstract* — **We live in the golden era of cosmology due to next redshift surveys will map a significant volume of the Universe across a wide range of redshifts. We analyze the clustering of the eBOSS luminous red galaxy sample (eBOSS LRG). We measure the multipole power spectra inferred from the NGC sample (107500 galaxies) and SGC sample (67316 galaxies) between redshifts 0.6 and 1.0, with effective redshift $z_{eff}=0.695$ and $z_{eff}=0.704$ for NGC and SGC samples, respectively. We use the non-linear perturbation theory to model the the redshift space multipoles of the galaxy power spectrum.**

## I. Introducción

Las estructuras a gran escala (LSS, por su siglas en inglés) del universo contienen información relevante para la astrofísica. Por ejemplo, información desde del universo temprano hasta restricciones de parámetros cosmológicos como la tasa de crecimiento de estructuras. Los catálogos de galaxias actuales como eBOSS observan el corrimiento al rojo, la huella característica de las galaxias, que permite situarnos en lo que se conoce como espacio del corrimiento al rojo. Sin embargo, tanto las velocidades peculiares de los objetos como la expansión misma del universo dejan una estructura más allá de la que existe en el espacio real. Estas características son conocidas como las distorsiones del corrimiento al rojo (RSD por sus siglas en inglés).

La evolución de las LSS a corrimientos al rojo altos y escalas grandes pueden ser modelados con la teoría lineal, mientras que el alcance de la teoría de perturbaciones puede ser extendida a escalas intermedias incluyendo expansiones del campo de densidad de materia hasta órdenes superiores.

En este trabajo, consideramos la teoría de perturbaciones estándar [1] (SPT, por sus siglas en inglés) para el estudio del espectro de potencia en el espacio del corrimiento al rojo de la encuesta de galaxia eBOSS. Consideramos dos técnicas adicionales: la teoría de campo efectiva (EFT, por sus siglas en inglés) y el esquema del resumado del infrarrojo (IR, por sus siglas en inglés).

Las encuestas observan diferente tipos de galaxias según su masa, luminosidad, tasa de formación estelar, entre otras características. La encuesta eBOSS observó galaxias llamadas luminosas rojas (LRG, por sus siglas en inglés). Las galaxias LRGs son caracterizadas por ser muy luminosas, razón por la cual pueden ser localizadas a corrimientos al rojo altos. Las LRGs son galaxias masivas, que contienen estrellas viejas con poca formación estelar en curso.

## II. Conjunto de datos de eBOSS

Se obtendrá el espectro de potencias del catálogo de galaxias de eBOSS. Describimos brevemente el conjunto de datos de la encuesta eBOSS. Las galaxias luminosas rojas fueron observados entre 0.6<z<1.0. El corrimiento al rojo permite situarnos en una escala de tiempo, z pequeños indican el universo más cercano.

La encuesta de galaxia de eBOSS forma parte de la colaboración Sloan Digital Sky Survey (SDSS), cuyo telescopio se encuentra en Nuevo México, Estados Unidos [2]. El catálogo de galaxias de eBOSS es observado a través de los dos hemisferios galácticos, referidos como la región galáctica del norte y del sur, las muestras NGC y SGC, respectivamente. La muestra de galaxias LRGs lleva una serie de procesos de limpieza con el objetivo de remover regiones con mala fotometría, galaxias que colisionan con espectros de cuásares, objetos que están muy cerca entre sí, entre otros efectos. Esta serie de limpieza remueve 17% de la huella del cielo inicial de eBOSS.

Las galaxias LRGs no son observadas cuando dos galaxias se encuentran muy cerca. Este efecto necesita tener asociado una función peso $w_{cp} = N_{targ}/N_{spect}$ donde $N_{targ}$ es el número de galaxias apuntados y $N_{spec}$ se refiere al número de galaxias con una observación espectroscópica. Similarmente ocurre con los objetos que no tienen una medición del corrimiento al rojo confiable. La función de peso asociada a este efecto es denotada por $w_{noz}$. De esta manera, la función colisión de peso es determinada por $w_{col} = w_{cp} \cdot w_{noz}$.

Otra función de peso necesaria para describir los efectos sistemáticos de observación y geométricos es $w_{sys}$. Esta función es necesaria ya que la densidad de objetos en el cielo, observados por eBOSS, no es contante. Finalmente, la función FKP, requerida para considerar la dependencia radial de la densidad, es definida como $w_{FKP}(z) = 1/[1 + n(z)P_0]$, donde $P_0$ es elegida como la amplitud del espectro de potencia P(k) a la escala de BAO, $k \sim 0.1 hMpc^{-1}$, $P_0 = 10000(h^{-1}Mpc)^3$.

Los objetos contenidos en el catálogo de galaxias LRGs tienen la siguiente función de peso total, la cual considera los 4 efectos descritos anteriormente,

$$w_{tot} = w_{FKP} \cdot w_{sys} \cdot w_{col}. \quad (1)$$

El corrimiento al rojo efectivo de la muestra está determinado por

$$z_{eff} = \left(\sum_{i>j} w_i w_j (z_i + z_j)/2\right) \bigg/ \left(\sum_{i>j} w_i w_j\right) \quad (2)$$

donde $w_i$ es el peso total de la galaxia i-ésima. Para la muestra NGC obtenemos $z_{eff} = 0.695$, y para la muestra SGC un corrimiento $z_{eff} = 0.704$. Estos valores fueron obtenidos al asumir que todos los pares de galaxias están separados por distancias entre 25 y 130 $h^{-1}Mpc$.

La región del cielo norte contiene 107500 galaxias, mientras que la muestra SGC contiene 67316 galaxias. La Fig. (1) muestra la huella del cielo observada por eBOSS en coordenadas RA y DEC. La distribución de las galaxias como función del corrimiento al rojo se muestra en la Fig. (2).

### A. Estimación del aglomeramiento de galaxias

Estimamos el aglomeramiento de galaxias en el espacio del corrimiento al rojo midiendo los multipolos del espectro de potencia. Asignamos las galaxias del catálogo de eBOSS y del catálogo de datos aleatorios a una malla regular Cartesiana, la cual permite usar el algoritmo basado en la transformada de Fourier (FT). El volumen de la encuesta total es distribuido en una caja cúbica de tamaño, y los dividimos en $N_g = 512^3$ celdas, cuya resolución y frecuencia Nyquist son $9.8 h^{-1}Mpc$ y $k_{Ny} = \pi N_g/L_b = 0.322 hMpc^{-1}$, respectivamente. Asignamos las partículas a la malla cúbica usando el esquema de interpolación TSC. Realizamos la medición del espectro de potencia con espaciados lineales con una anchura de $\Delta k = 0.01 hMpc^{-1}$ hasta $k_{max} = 0.32 hMpc^{-1}$.

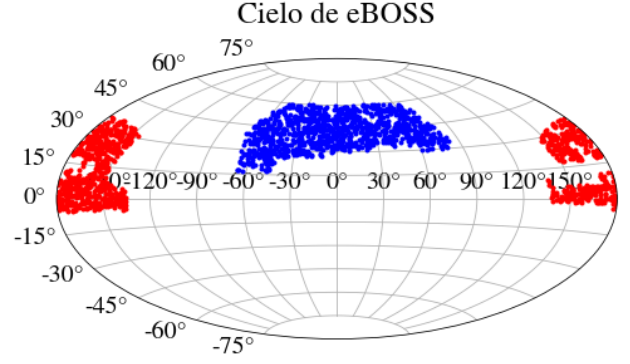

Fig. 1. Distribución del cielo de las galaxias en la encuesta de galaxias eBOSS usando la proyección Aitoff. Los puntos azules denotan la región NGC, y los rojos la región SGC. Por visibilidad, sólo 1000 objetos de cada región son usados.

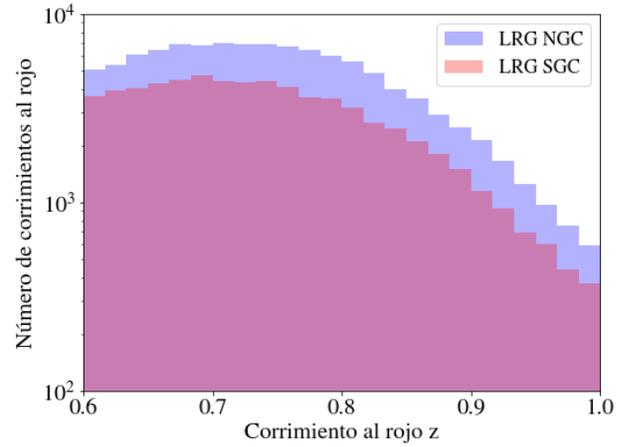

Fig. 2. Histograma del corrimiento al rojo de las regiones norte (NGC) y sur (SGC) usados en el análisis de las estructuras a gran escala de eBOSS.

### B. Estimador del espectro de potencia

Para calcular los multipolos del espectro de potencia, iniciamos definiendo la función de campo de densidad de peso $F(\mathbb{r}_i)$ definido como la diferencia entre la densidad de galaxias y la densidad del catálogo sintético de objetos aleatorios multiplicado por una función de peso $w_{tot}(\mathbb{r}_i)$, dividido sobre un factor de normalización, es decir,

$$F(\mathbb{r}_i) = \frac{w_{tot}(\mathbb{r}_i)[n_{gal}(\mathbb{r}_i) - \alpha_{ran} n_{ran}(\mathbb{r}_i)]}{\int d\mathbb{r} [n_{gal} w_{tot}(\mathbb{r})]^2} \quad (3)$$

donde $n_{gal} = \sum_i \delta(\mathbb{r} - \mathbb{r}_i)$, con $\mathbb{r}_i$ la localización de la i-ésima galaxia, y $\alpha_{ran}$ es la razón entre el número de galaxias y objetos aleatorios con su peso asociado. El factor de normalización de la Ec. (3) es introducido debido a que el catálogo de objetos aleatorios tiene una mayor densidad que el conjunto de datos de galaxias. En este trabajo, utilizamos catálogos de datos aleatorios con 20 veces más objetos que los catálogos de galaxias.

El estimador para los multipolos del espectro de potencia es

$$P^{(\ell)}(k) = (2\ell+1)\int \frac{d\Omega_k}{4\pi}\int d\mathbb{r}_1\, F(\mathbb{r}_1)e^{-i\mathbb{k}\cdot\mathbb{r}_1}$$
$$\times \int d\mathbb{r}_2\, F(\mathbb{r}_2)e^{-i\mathbb{k}\cdot\mathbb{r}_2}\mathcal{L}\left(\frac{\mathbb{k}\cdot\mathbb{r}_h}{|\mathbb{k}||\mathbb{r}_h|}\right) \quad (4)$$

donde $\mathbb{r}_h = (r_1+r_2)/2$, $\mathcal{L}_\ell$ son los polinomios de Legendre de orden $\ell$.

### III. MODELO DEL ESPECTRO DE POTENCIAS

Para modelar los multipolos del espectro de potencias del catálogo de galaxias usaremos las técnicas perturbativas, el resumado del infrarrojo y la teoría de campo efectiva. La teoría de perturbaciones estándar consiste en expandir los campos de densidad y de velocidad hasta tercer orden (1-loop, SPT por sus siglas en inglés), la teoría efectiva agrega contratérminos necesarios para la dinámica de escalas pequeñas y el resumado IR corrige las modulaciones de las oscilaciones acústicas de bariones en el espectro. Se considera la expansión del sesgo hasta 1-loop para la predicción teórica del espectro de galaxias.

En el espacio del corrimiento al rojo, el espectro de potencia es caracterizado por el coseno del ángulo entre la línea de visión y el vector de onda de un modo dado $\mathbb{k}$, $\mu = (\mathbb{z}\cdot\mathbb{k})/|\mathbb{k}|$. El espectro de potencias en el espacio del corrimiento al rojo tiene la siguiente estructura

$$P_{gg,RSD}(z,k,\mu) = P_{K,RSD}(z,k,\mu)$$
$$+P_{1-loop,SPT,RSD}(z,k,\mu) + P_{ctr,RSD}(z,k,\mu)$$
$$+P_{\epsilon\epsilon,RSD}(z,k,\mu) \quad (5)$$

donde el primer término de la Ec. (5) hace referencia al orden lineal en el espacio del corrimiento al rojo llamado el término de Kaiser [3], el segundo término a la siguiente corrección, conocida como de un bucle (1-loop) [1], el tercer término al contratérmino proveniente de la teoría efectiva [4, 5] y el último término proviene de las contribuciones estocásticas. Para una mayor discusión sobre la Ec. (5) vea [6,7,8]. Usualmente, se engloba la información angular en los primeros multipolos, de tal forma que

$$P_{gg,RSD}(z,k,\mu) = \sum_{\ell par}^{\ell=4} \mathcal{L}_\ell(\mu)P_\ell(z,k). \quad (6)$$

Los flujos de las estructuras a gran escala generan una huella característica en las escalas de las oscilaciones acústicas de bariones (BAO), por lo que es necesario usar una técnica adicional llamada resumado del infrarrojo (IR), la cual permite resumar las contribuciones del infrarrojo [9]

$$P_{gg}^{IR}(z,k,\mu) = (b_1(z)+f(z)\mu^2)^2$$
$$\times \left(P_{nw}(z,k) + e^{-k^2\Sigma_{tot}^2(z,\mu)}P_w(z,k)\left(1+k^2\Sigma_{tot}^2(z,\mu)\right)\right)$$
$$+P_{gg,nw,RSD,EFT}(z,k,\mu)$$
$$+e^{-k^2\Sigma_{tot}^2(z,\mu)}P_{gg,w,RSD,EFT}(z,k,\mu) \quad (7)$$

con

$$P_{gg,RSD,EFT}(z,k,\mu) = P_{gg,RSD,1-loop,SPT}(z,k,\mu)$$
$$+P_{ctr,RSD}(z,k,\mu) \quad (8)$$

donde la función de amortiguamiento total está definida como

$$\Sigma_{tot}^2(z,\mu) = (1+f(z)\mu^2)\big((2+f(z))\big)\Sigma^2(z)$$
$$+f^2(z)\mu^2(\mu^2-1)\delta\Sigma^2(z),$$

y $\Sigma^2(z)$ es la función de amortiguamiento en el espacio real

$$\Sigma^2(z) \equiv \frac{1}{6\pi^2}\int_0^{k_s} dq\, P_{nw}(z,q)\left[1-j_0\left(\frac{q}{k_{osc}}\right)+2j_2\left(\frac{q}{k_{osc}}\right)\right].$$

El número de onda correspondiente a la longitud de onda de BAO ($\ell_{osc} 110 h^{-1} Mpc$) es $k_{osc}$, y la nueva contribución en el espacio al corrimiento al rojo es

$$\delta\Sigma^2(z) \equiv \frac{1}{2\pi^2}\int_0^{k_s} dq\, P_{nw}(z,q) j_2\left(\frac{q}{k_{osc}}\right).$$

Usamos las técnicas mencionadas anteriormente para modelar los multipolos del espectro de potencia. En la Fig. (3) se muestran las mediciones de los multipolos para las dos muestras, NGC y SGC. En la Fig. (4) se muestran el modelo teórico y los datos obtenidos por las encuestas espectroscópicas.

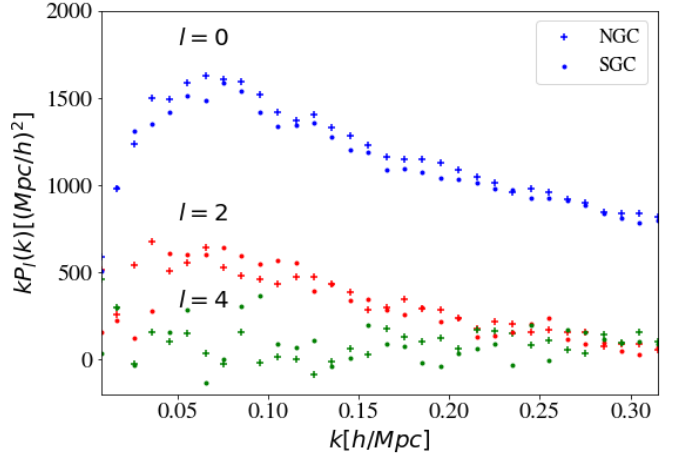

Fig. 3. Multipolos del espectro de potencia para las regiones norte (NGC) y sur (SGC).

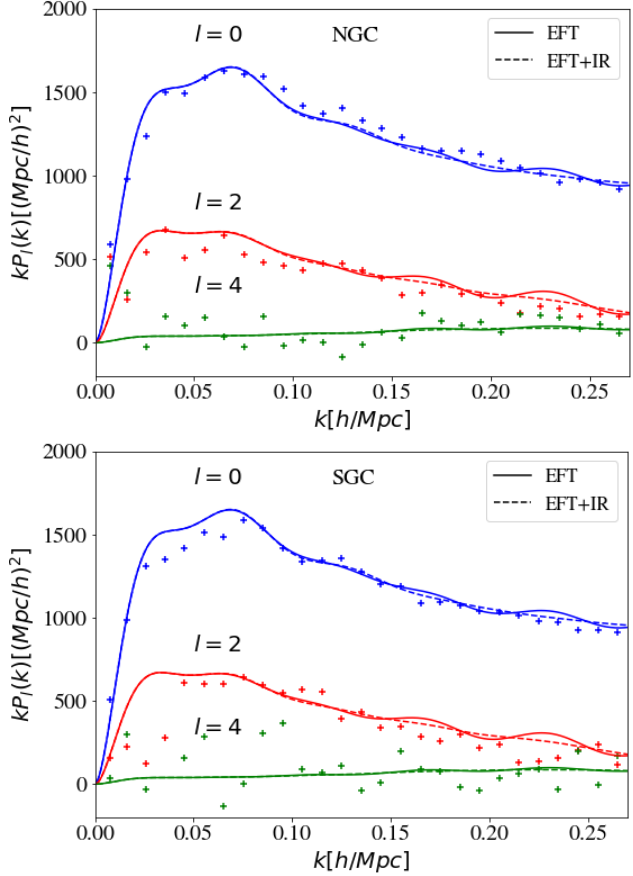

Fig. 4. Multipolos del espectro de potencia para la a) región norte (NGC) y b) el cielo sur (SGC). Las líneas sólidas indican los multipolos considerando la teoría efectiva, mientras que las lineas punteadas considerando EFT y la técnica del resumado al infrarrojo (IR).

## IV. Discusión

Si bien los resultados de los datos de las regiones NGC y SGC presentan demasiada dispersión debido a que hay pocos datos en los catálogos, la combinación de estos con las encuestas previas como BOSS dan una cantidad de 370000 galaxias LRGs que mejorarán los resultados. En un futuro, se pretende trabajar con esta combinación de datos y comparar los resultados con las técnicas perturbativas revisadas en este trabajo. En este trabajo, solo trabajamos con los catálogos NGC y SGC, por separado, para modelar sus estadísticas de dos puntos con teoría de perturbaciones no lineal.

## V. Conclusión

Con base en la teoría de perturbaciones estándar en este trabajo calculamos los multipolos del espectro de potencias de galaxias de la encuesta eBOSS. Notamos que la teoría de perturbaciones describe muy bien los multipolos hasta $k \sim 0.3 hMpc^{-1}$. Se incluyeron los efectos del sesgo no lineal y las distorsiones del espacio del corrimiento al rojo, el resumado del infrarrojo, los contratérminos provenientes de la teoría de campo efectiva. Los contratérminos son necesarios para modelar la física de pequeña escala que está fuera del alcance de la teoría de perturbaciones estándar. Mientras que el resumado del infrarrojo sirve para modelar los flujos de los grandes bultos que afectan las oscilaciones de BAO.

## Referencias

[1] F. Bernardeau, S. Colombi, E. Gaztanaga y R. Scoccimarro, "Large scale structure of the universe and cosmological perturbation theory", Phys. Rep. 367, 1 (2002).
[2] Collaboration eBOSS, "The Completed SDSS-IV extended Baryon Oscillation Spectroscopic Survey: Large-scale structure catalogues for cosmological analysis", Monthly Notices of the Royal Astronomical Society, vol 498 (2020).
[3] N. Kaiser, "Clustering in real space and redshift space", Mon. Not. R. Astron. Soc. 227, 1 (1987).
[4] D. Baumman, A. Nicolis, L. Senatore, y M. Zaldarriaga, "Cosmological nonlinearities as an effective fluid", J. Cosmol. Astropart. Phys. 07 (2012) 051.
[5] J.J.M. Carrasco, M.P. Hertzberg, y L. Senatore, "The effective field theory of cosmological large scale structures", J. High Energy Phys. 09 (2012) 082.
[6] D. V. Gomez-Navarro, "Estudiando las estructuras a gran escala del Universo con la teoría de perturbaciones no lineal", En preparación para las memorias de la RNAFM (2022).
[7] L. Senatore y M. Zaldarriaga, "Redshift space distortions in the effective field theory of large scale structures", arXiv:1409.1225.
[8] A. Perko, L. Senatore, E. Jennings, y R.H. Wechsler, "Biased tracers in redshift space in the EFT of large-scale structure", arXiv:1610.09321.
[9] Tobias Baldauf, Mehrdad Mirbabayi, Marko Simonović, and Matias Zaldarriaga Phys. Rev. D **92**, 043514 (2015).